\documentclass[aps,prl,reprint,showpacs,showkeys]{revtex4-1}
\usepackage{microtype}
\usepackage{amsthm}
\usepackage{amsmath}
\usepackage{amsfonts}
\usepackage[mathscr]{eucal}
\usepackage{amssymb,epsfig,setspace}
\usepackage{graphicx}
\usepackage{caption}
\usepackage{subcaption}
\begin{document}
\title[Spectrum of the Rabi model]{Full spectrum of the Rabi model}%
\author{Andrzej J.~Maciejewski} \email{maciejka@astro.ia.uz.zgora.pl}
\affiliation{J.~Kepler Institute of Astronomy, University of Zielona
  G\'ora, Licealna 9, PL-65--417 Zielona G\'ora, Poland.}%
\author{Maria Przybylska}%
\email{M.Przybylska@proton.if.uz.zgora.pl} \affiliation{ Institute of
  Physics, University of Zielona G\'ora, Licealna 9, PL-65--417 Zielona
  G\'ora, Poland }%
\author{Tomasz Stachowiak} \email{stachowiak@cft.edu.pl}
\affiliation{%
  Center for Theoretical Physics PAS, Al. Lotnikow 32/46, PL-02-668
  Warsaw, Poland }%

\date{\today}%

\begin{abstract}
  It is shown that in the Rabi model, for an integer value of the spectral
  parameter $x$,
  in addition to the finite number of the classical Judd states there
  exist infinitely many possible eigenstates. These eigenstates exist
  if the parameters of the problem are zeros of a certain
  transcendental function; in other words, there are infinitely many
  possible choices of parameters for which integer $x$ belongs to the
  spectrum. Morover, it is shown that the classical Judd eigenstates
  appear as degenerate cases of the confluent Heun function.
\end{abstract}
                      
\keywords{Rabi model; quantum optics; Bargmann--Fock representation\\[2ex]
${}^{\dagger}$ Corresponding author. E-mail address: M.Przybylska@proton.if.uz.zgora.pl (M. Przybylska); tel +48 68 328 2914; fax   +48 68 328 2920
 }%
\maketitle


The aim of this paper is to specify the full spectrum of the classical
Rabi system and to describe it in a uniform framework. Although the
Rabi model has been investigated for over seventy years, its complete
spectrum has not been determined, and some of its important properties
are unknown.

The Rabi model was introduced by Rabi~\cite{Rabi:36::} to describe the
action of a rapidly-varying, weak magnetic field on an oriented atom
possessing nuclear spin. Now it is one of the ubiquitous models
describing various quantum systems, including cavity and circuit
quantum electrodynamics (QED), quantum dots, polaronic physics and
trapped ions~\cite{Niemczyk:10::,Englund:07::,FornDiaz:10::,Crepsi:12::,Casanova:10::,Gunter:09::}. Coupling between atoms and radiation
is usually very weak. In such cases, the rotating wave approximation
(RWA) is well-justified in the Rabi model, and it is called the
Jaynes--Cummings model. However, recent achievements in circuit QED
have enabled the exploration of regimes---e.g. the ultra-strong and
the deep strong coupling regimes of light-atom interaction---in which
the Jaynes-Cummings model begins to fail and the observed phenomena
can be explained only by the full Rabi system. More importantly, there is now evidence
that even for weak couplings the RWA predicts behavior that differs
significantly from
that of the full system \cite{Larson:12::,Schiro:12::}. For this reason, the
Rabi model has now been revisited by many authors with the intent of a
detailed analysis of its spectrum. To be more specific, the
Hamiltonian in question is
\begin{equation}
  H = a^{\dag}a + \mu\sigma_z + 
  \lambda(\sigma_{+}+\sigma_{-})(a^{\dag}+a),
\end{equation}
where $a$, $a^{\dag}$ are the photon annihilation and creation
operators satisfying the canonical commutation relation
$[a,a^{\dag}]=1$ which arise due to the harmonic oscillator representation of
quantized electromagnetic field; $\mu$, $\lambda$ are the level separation and
photon-atom coupling constants, and the $\sigma$ spin
operators are here taken to be
\begin{equation}
\sigma_{+} = \begin{pmatrix} 0 & 0\\ 1 & 0\end{pmatrix},\;\;
\sigma_{-} = \begin{pmatrix} 0 & 1\\0&0\end{pmatrix},\;\;
\sigma_z = \begin{pmatrix} 1 & 0\\0&-1\end{pmatrix}.
\end{equation}
The RWA is obtained by neglecting $\sigma_- a$ and $\sigma_+
a^{\dag}$.

The Bargmann--Fock Hilbert space $\mathscr{H}$, which we will be using, is a
space of entire functions of one complex variable $z\in\mathbb{C}$ with a
scalar product given by \[
\langle f,g\rangle:=
\dfrac{1}{\pi}\int_\mathbb{C}\overline{f(z)}g(z)e^{-|z|^2}\mathrm{d}
(\Re(z))\mathrm{d}(\Im(z)).
\]

In this Hilbert space, the operators $a^{\dag}$ and $a$ become
multiplication and differentiation with respect to a complex variable
$z$. This stems from the commutation relation
$[\partial_z,z]=1$. Thus, the Rabi model in this representation is
described by the following system of two differential equations
\begin{equation}
  \begin{split}
    (z+\lambda)\dfrac{\mathrm{d}\psi_1}{\mathrm{d}z}=&(E-\lambda
    z)\psi_1-\mu\psi_2,\\
    (z-\lambda)\dfrac{\mathrm{d}\psi_2}{\mathrm{d}z}=&(E+\lambda
    z)\psi_2-\mu\psi_1,
  \end{split}
  \label{eq:syst}
\end{equation}
where $E$ is the energy, for details see \cite{Kus:86::}. In this
representation, the two component wave function $\psi=(\psi_1,
\psi_2)$ is an element of a Hilbert space
$\mathscr{H}^2=\mathscr{H}\times\mathscr{H}$. An entire function
$f(z)$ belongs to $\mathscr{H}$ if it has proper growth at
infinity see, e.g.,~\cite{Bargmann:61::,Vourdas:06::}. Namely their growth order
must be less or equal 2, and if it is equal 2, then it must be of the type less than 1/2. Additionally, an entire function of growth order 2 and type 1/2 can be normalizable but in this case an separate analysis is necessary. For
definition of the growth order and type see, e.g.,~\cite{Levin:96::}. One can estimate from the above the growth order of entire solutions of a differential equation (or a system of differential equations) with rational coefficients by the degree at infinity of these coefficients, see~\cite{Laine:93::}. These estimations
for the equations describing the Rabi model show that all entire solutions of these equations have the growth order less or equal to one, so they are normalizable.

Let us remark here that in most cases when the
Bargmann-Fock representation is used, it is not checked if obtained
eigenfunctions are normalizable. We call attention to the fact that the
Bargmann-Fock representation
has several peculiar properties. For example, from the
fact that an entire function $f(z)$ is an element of $\mathscr{H}$,  it
does not follow that entire function $g(z):=zf(z)$ belongs
to  $\mathscr{H}$.

The energy spectrum of the Rabi system has been investigated by many authors
and various approaches are used. The most popular ones are numerical
techniques equivalent to a large-scale diagonalization of a suitably
defined finite subspace of the full Hilbert space
\cite{Graham:84::,Durst:86::,Feranchuk:96::,Tur:01::,Emary:02::a}. Thus,
the eigenvalues and eigenstates determined in this way are inaccurate
by construction. Moreover, from such results it is difficult to
extract the complete physical description of the system.

Schweber was the first to apply the Bargmann space to the Rabi model
\cite{Schweber:67::}, and the first known analytical results are due to Swain
\cite{Swain:73::,Swain:73::a},
who employs a continued fraction technique to calculate transition
amplitudes for spontaneous absorption and emission, and from this
he tries to obtain the spectrum and the eigenfunctions of the Rabi
Hamiltonian. However, this approach is ineffective. Reik and others
\cite{Reik:86::} adapt Judd's method \cite{Judd:79::}, used originally
for the Jahn-Teller system. Whereas Judd uses a power series
substitution, Reik and others apply a Neumann series. They show that
if $x:=E+\lambda^2$ is an integer and parameters of the system satisfy
some equalities, then the Neumann series terminates, giving isolated,
exact solutions known as the Judd solutions. These conditions were
also obtained by Ku{\'s} \cite{Kus:86::} using another substitution.
As described in~\cite{Kus:86::}, if, for a given $n\in\mathbb{N}$,
parameters of the problem $(\lambda,\mu)$ belong to the following
algebraic set
\begin{equation}
  \label{eq:Jn}
  \mathscr{J}_{n}:= \left\{ (\lambda,\mu)\in\mathbb{R}^2\,|\, 
    P_n(\lambda^2,\mu^2)=0 \right\},
\end{equation} 
where $P_n(\lambda^2,\mu^2)$ is a certain polynomial of degree $n$,
then $x=n$ is a doubly degenerate eigenvalue of the problem.
Geometrically speaking, for each $n\in\mathbb{N}$, Eq.~\eqref{eq:Jn}
defines $n$ algebraic ovals in the $(\lambda,\mu)$-plane. The two
corresponding eigenstates are known and given in terms of elementary
functions \cite{Kus:85::}. These eigenstates have also been expressed
explicitly by means of the Neumann series expansion
\cite{Reik:86::}. Judd's solutions can be recognized as quasi-exact
solutions \cite{Koc:02::} in that only finitely many states are
expressible in terms of elementary functions when appropriate
conditions are met. In later work, it is shown \cite{Braak:11::} that
for $x\not\in\mathbb{N}$, the spectrum of the Rabi problem can be
determined as zeros of certain transcendental functions
$G_{\pm}(x,p)=0$. For any pair $p=(\lambda,\mu)$ of parameters there
exist discrete values of admissible energy.  Moroz \cite{Moroz:12::a}
proposes yet another method of determining the spectrum.

\begin{figure*}[t]
  \centering
  \begin{subfigure}[b]{0.33\textwidth}
    \centering
    \includegraphics[width=\textwidth]{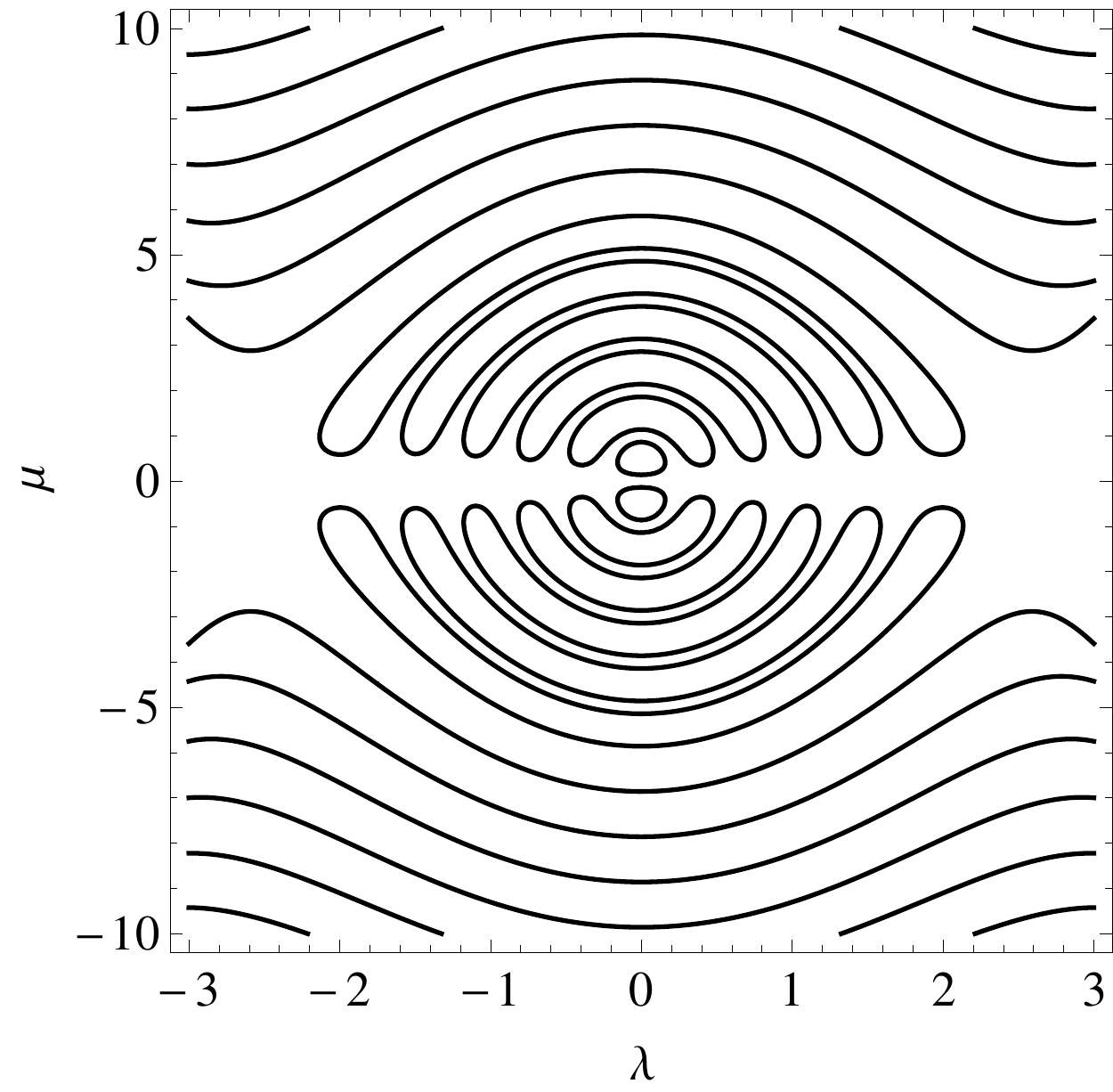}
    \caption{\label{fig:pii2g} }
  \end{subfigure}%
  \begin{subfigure}[b]{0.33\textwidth}
    \centering
    \includegraphics[width=\textwidth]{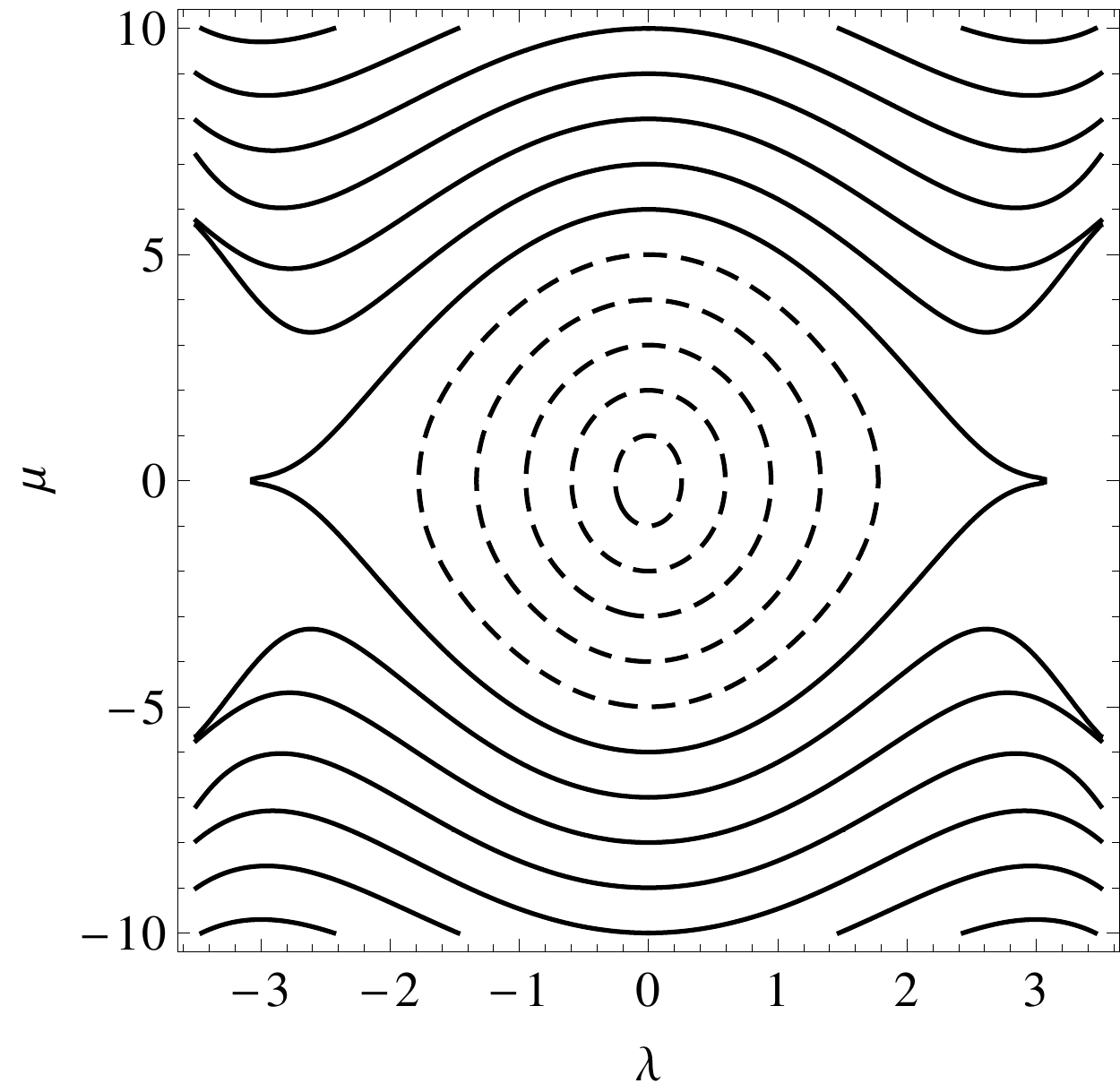}
    \caption{\label{fig:3}}
  \end{subfigure}
  \begin{subfigure}[b]{0.33\textwidth}
    \centering
    \includegraphics[width=\textwidth]{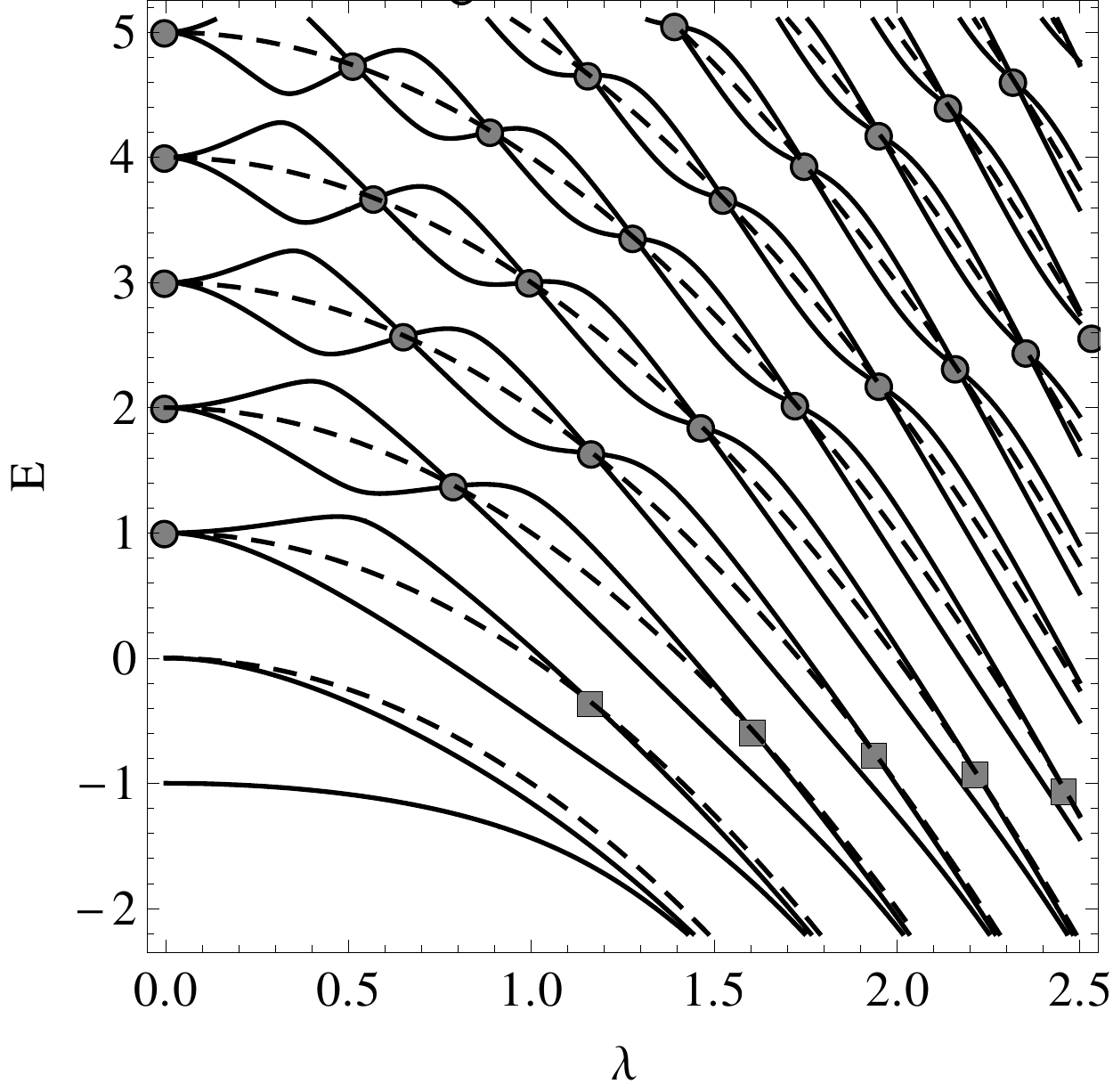}
    \caption{\label{fig:spectcom2} }
    \label{fig:mouse}
  \end{subfigure}
  \caption{a) Curves $ \mathscr{S}_{x}$ for $x=2+\pi$ on
    $(\lambda,\mu)$-plane. b) Curves $\mathscr{F}_5$ (continuous
    lines), $\mathscr{J}_5$ (dashed lines) in the $(\lambda,\mu)$
    plane. c) Energy spectrum for resonant case $\mu=1$. Gray circles
    are Juddian points and gray squares represent new elements of the
    spectrum. Energy baselines $E+\lambda^2=p$ with $p\in\mathbb{N}_0$
    are plotted with dashed lines.}
\end{figure*}

Following Braak's work \cite{Braak:11::}, the Rabi model was deemed
completely solved \cite{Physics.4.68}. In other words, consensus held
that all the spectral values of $x$ were known, as well as the
corresponding eigenvalues. However, it is demonstrated in this Letter
that this conclusion  is not completely true. To advance the argument, it is pertinent to establish what \emph{complete solution} means. For
arbitrary fixed values of parameters $p=(\lambda,\mu)\in\mathbb{R}^2$,
all values of energy $E$ (or parameter $x$) for which
system~\eqref{eq:syst} has a non-zero entire solution
$\left(\psi_1(z;x,p),\psi_2(z;x,p)\right)$ must be specified. Thus, in
three dimensional parameters space $\mathscr{P}\subset\mathbb{R}^{3}$
those points $(x,\lambda,\mu)\in \mathscr{P}$ for which
system~\eqref{eq:syst} has an entire solution must be
distinguished. All such points form a complicated set $\mathscr{S}$,
which is called the spectral set. To represent it graphically it is
cut with a plane $\mu=\mu_0$ or a plane $x=x_0$. For a full
description of $\mathscr{S}$ both types of intersections are
required. Subsets of $\mathscr{S}$ with fixed values $x=x_0$ are
denoted by $\mathscr{S}_{x_{0}}$. For a generic value of $x$, the set
$\mathscr{S}_{x}$ contains infinitely many curves. An example for
$x=2+\pi$ is shown in Fig.~\ref{fig:pii2g}.

Maciejewski et al. \cite{Maciejewski:12::x} show that if $x$ is not
an integer, then the spectrum can be defined by one closed-form
condition expressed as the Wronskian of the confluent Heun functions
$W(x,\lambda,\mu)=0$. As an alternative description, a large part of
spectral set $\mathscr{S} $ is given by
\begin{equation}
  \label{eq:6}
  \mathscr{W}:=
  \left\{ (x,\lambda,\mu)\in \mathscr{P}\, |\, W(x,\lambda,\mu)=0  \right\}
  \subset \mathscr{S}\text{.}
\end{equation}
The Heun function generalizes the Gauss hypergeometric, wave
spheroidal, Lam\'e and Mathieu functions, and finds numerous
applications in quantum physics \cite{Slavyanov:00::,Hortacsu:11::}. Since the
general solution is expressible by this function, so are also the Juddian ones,
which correspond to its degenerate cases. This fact was used in
\cite{Zhong:13::} to obtain all parts of the spectrum, using uniform formulae
for the solutions regardless of the value of $x$, and the authors use both a
corrected version of Braak's method (checking more than one $z$ value for the
$G_{\pm}$ functions) and Wronskians, both of which elements can already be found
in \cite{Maciejewski:12::x}

It is now shown that $\mathscr{W} $ contains all points
$(x,\lambda,\mu)$ of the spectral set $\mathscr{S} $ with non-integer
$x$. This follows from considerations in earlier work
\cite{Maciejewski:12::x}. The crux of the problem is the determination
of points of the spectral set with $x \in \mathbb{Z}$. Such values of
$x$ are not considered in other recent work
\cite{Braak:11::}. Clearly, Judd states give such points, i.e.,
$\mathscr{J}_n\subset\mathscr{S}_n$. Hence, the question is, if, apart
from Judd states, there exist states with integer values of spectral
parameter $x$. This Letter argues that they do exist. For a fixed
integer value of spectral parameter $x=n$, the set $\mathscr{S}_n$ has
two components. The first one, $\mathscr{J}_n$, consists of a finite
number of curves corresponding to classical Judd states. The second
one, $\mathscr{F}_n$, consists of infinitely many curves,
corresponding to states that have not received attention in prior
works. We can provisionally define it as $\mathscr{F}_n = \mathscr{S}_n
\setminus \mathscr{J}_n$ (with a more specific definition to follow),
and one goal of the remainder of the Letter will be to understand its
properties.
A graphical example is given in Fig.~\ref{fig:3}, where ovals
$\mathscr{J}_{5}$ on the $(\lambda,\mu)$-plane are drawn with dashed
lines, and curves $\mathscr{F}_5$ are drawn with continuous lines. 
That there are infinitely many branches of $\mathscr{F}_n$  can be intuitively seen from the fact, that the curves
go continuously through the $\lambda=0$ line, and in that case
the solution of the system~\eqref{eq:syst} is $\psi_1=c_1z^{E-\mu}+c_2z^{E+\mu}$, 
$\psi_2=c_1z^{E-\mu}-c_2z^{E+\mu}$. These are entire functions only
for non-negative integer values of the exponents, so for a baseline $E=n$, they give countably infinitely many
possible values of $\mu$ where $\mathscr{F}_n$ crosses the $\lambda=0$ axis.

\begin{figure}[htpp]
  \begin{center}
    \includegraphics[width=0.482\textwidth]{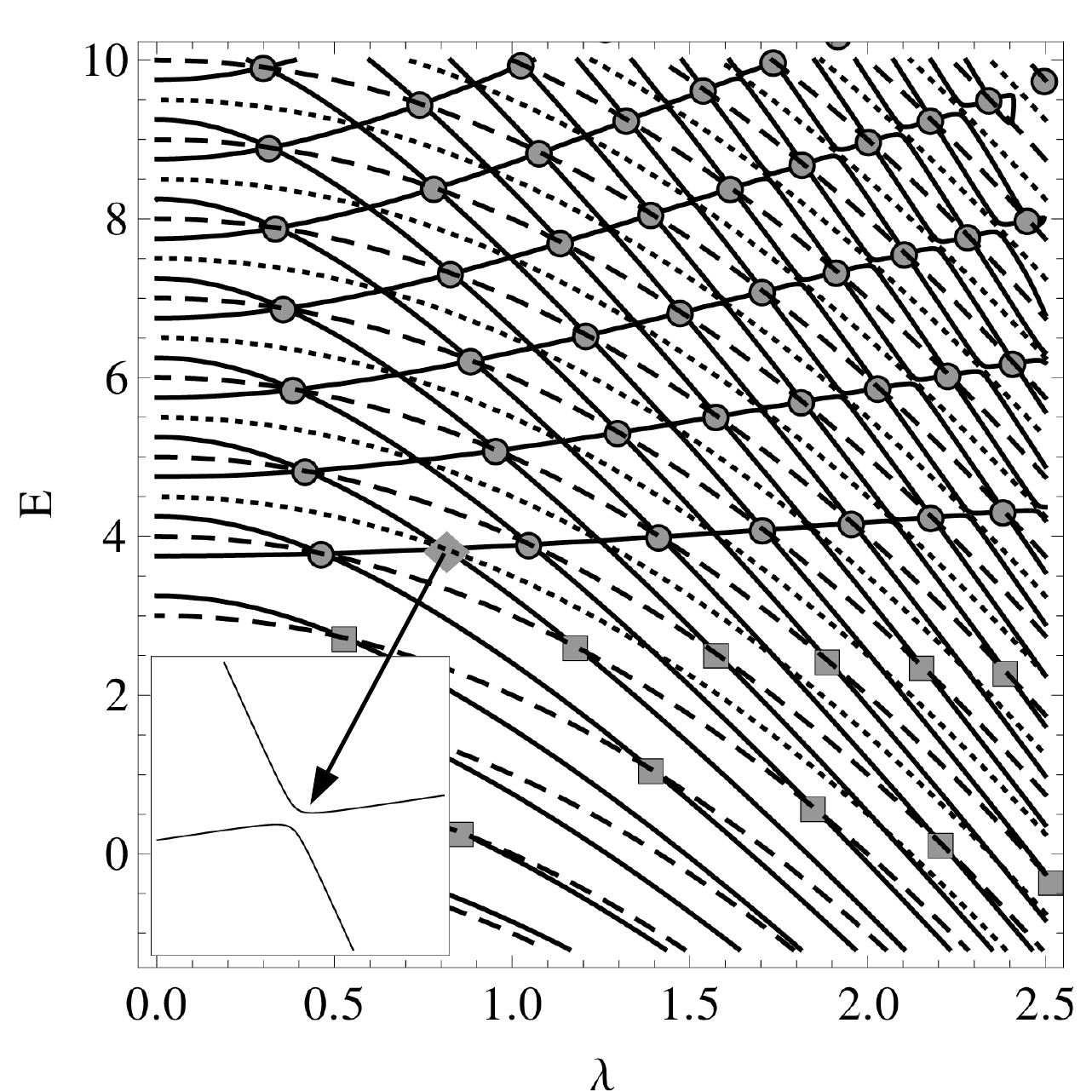}
  \end{center}
  \caption{\label{fig:spectcom23} Energy spectrum for
    $\mu=3\frac{3}{4}$. An apparent level crossing is marked with a
    diamond, shown on magnification of the corner. The range of
    variables in this inset is $\lambda\in[0.806, 0.817]$ and
    $E\in[3.835, 3.850]$.}
\end{figure}

The spectrum of the model for $\mu=1$ is shown in
Fig.~\ref{fig:spectcom2}. The new elements of spectrum, denoted by
gray squares, lie on the energy baselines $E+\lambda^2=n$, where $n$
is an integer, much like the Judd eigenstates. However, these new
elements are not degenerated. Another phenomenon is apparent in
Fig.~\ref{fig:spectcom23}, which shows the spectrum for
$\mu=3\frac{3}{4}$. Also plotted, with dotted lines, are the energy
baselines corresponding to half-integer values of $n$. Manifestly, for
each line of the spectrum with $E>3$, for half-integer energy baseline
between pairs of Judd states, the levels of the same parity seem to
intersect. However, a magnification of the bottom-left corner of
Fig.~\ref{fig:spectcom23} shows that at these points the curves do not
cross.

As aforementioned, for determination of the spectrum, the problem is
reduced to the purely mathematical question of assessing if, for a
given $(x,\lambda,\mu)$, system~\eqref{eq:syst} admits an entire
solution. The approach taken to this problem here is based on the
classical analytic theory of complex differential equations (see Ince
\cite{Ince:44::}). Although presented for the Rabi model alone, it is
applicable to a wide class of systems.

It is convenient to transform system~\eqref{eq:syst} to one
second-order equation, putting $z=\lambda(2y-1)$ and
$\psi_1(y)=\exp(2\lambda^2 y)v(y) $. Eliminating $\psi_2$ from
system~\eqref{eq:syst} the confluent Heun equation is obtained
\cite{Ronveaux:95::},
\begin{equation}
  v''+\left(\alpha+\dfrac{\beta+1}{y}+\dfrac{\gamma+1}{y-1}
  \right)v'+\left(\dfrac{ \theta }{
      y}+\dfrac{\xi}{y-1}\right)v=0,
  \label{eq:Heun}
\end{equation} 
with parameters $\alpha=4\lambda^2$, $\beta=-x$, $\gamma=-1-x$, and
\begin{equation}
  \theta= 4 \lambda^2 + \mu^2-x^2,\quad \xi=x(x- 4 \lambda^2) -
  \mu^2.
\end{equation}  
Instead of $\theta$ and $\xi$, the parameter used are
$\delta:=2\lambda^2$, and
\begin{equation}
  \eta:=\frac{1}{2} \left( 1+x+x^2\right)-\mu^2-2\lambda^2 \left(x+1\right)\text{.}
  \label{eq:par_heun_def}
\end{equation} 
Eq.~\eqref{eq:Heun} has two regular singular points at $y=0$, with
exponents $\left(\rho_1^{(0)},\rho_2^{(0)}\right)=(x,0)$, and at
$y=1$, with exponents
$\left(\rho_1^{(1)},\rho^{(1)}_{2}\right)=(x+1,0)$. Infinity is an
irregular singular point.

From the general theory (see Whittaker ~\cite[Ch.~X
in][]{Whittaker:35::}), it is known that at each regular singular
point $y_{\star}\in\{0,1\}$ there are two independent local
solutions. If the difference of exponents
$\left(\rho^{\star}_1-\rho^{\star}_2\right)$ is not an integer, then
these solutions are of the form
\begin{equation}
  \label{eq:1}
  v_i^{\star}(y)=(y-y_{\star})^{\rho_i^{\star}}h_i^{\star}(y), \qquad i=1,2, 
\end{equation} 
where $h_i^{\star}(y)$ are holomorphic at $y_{\star}$.

Initially it is assumed that $x$ is not a non-negative integer and
that it belongs to the spectrum. Then the corresponding eigenvector is
given by an entire solution $v(y)$ of~\eqref{eq:Heun}. An expansion of
$v(y)$ at a singular point $y_{\star}$, coincides, up to a
multiplicative constant, with a local solution holomorphic at
$y_{\star}$. In the considered case, the only locally holomorphic
solutions are those corresponding to exponents
$\rho_2^{(0)}=\rho_2^{(1)}= 0$. Thus, solutions $v^{(0)}_2(y)$ and
$v^{(1)}_2(y)$ must coincide in the common part of their domains of
definitions as they are expansions of one entire solution
$v(y)$. Hence the Wronskian of these solutions must vanish. This gives
a restriction on the parameters of the problem, and, in effect, it
allows for the determination of the spectrum of the problem. However,
expansions $v^{(0)}_2(y)$ and $v^{(1)}_2(y)$ are taken around
different points. This can be dealt with by exploiting the fact that
both these solutions can be expressed in terms of confluent Heun
functions \cite{Ronveaux:95::}.  Denoting
\begin{equation*}
  \label{eq:hc}
  H_0(y):=\operatorname{HeunC}(a_0
  ; y), \quad  H_1(y):=\operatorname{HeunC}(a_1;1- y), 
\end{equation*}
the confluent Heun functions \cite{Ronveaux:95::}, with parameters
$a_0:=(\alpha,\beta,\gamma,\delta,\eta) $ and
$a_1:=(-\alpha,\gamma,\beta,-\delta,\delta+\eta) $, it is found that
$v^{(0)}_2(y)=H_0(y)$ and $v^{(1)}_2(y)=H_1(y)$. Thus the Wronskian of
solutions is given by
\begin{equation}
  \label{eq:w}
  w(x,p;y):= H_1(y)H_2'(y) - H_1'(y)H_2(y).
\end{equation}
The Wronskian shown in \eqref{eq:w} must vanish in the intersection of
domains of solutions. But if it vanishes at one point of this
intersection, then it vanishes identically. Taking $y=1/2$ and the set
$W(x,p):=w(x,p;1/2)$, the zeros of $W(x,p)$ then determine the
spectrum of the problem with the assumption that $x$ is not a
non-negative integer. This explains why $\mathscr{W}$, given
by~\eqref{eq:6}, is a subset of the spectral set $\mathscr{S}$.

If $x$ is a non-negative integer, then the problem is more complex. In
such a case, generally, only local solutions corresponding to
exponents $\rho_{1}^{(i)}$ with $i=0,1$ have the form \eqref{eq:1},
i.e. they are locally holomorphic. Solutions corresponding to
exponents $\rho_{2}^{(0)}=\rho_{2}^{(1)}=0$ may have logarithmic
terms, and such solutions have the following form
\begin{equation}
  \label{eq:2}
  v_2^{(i)}(y)=r_i v_1^{(i)}(y)\ln(y-y_{i}) +g_i(y), \qquad i=0,1,
\end{equation}  
where $r_i$ are constants depending on parameters, $y_0=0$, $y_1=1$,
and $g_i(y)$ are holomorphic at $y_i$, for $i=0,1$.

If $x$ is a non-negative integer that belongs to the spectrum, the
corresponding eigenvector is given by the entire solution $v(y)$
of~\eqref{eq:Heun}. If local expansions of $v(y)$ at $y_i$ do not
vanish at $y_i$, then they have to coincide, up to a multiplicative
constant, with local solutions of $v_2^{(i)}(y)$ with $r_1=r_2=0$. By
the Frobenius method \cite{Ince:44::}, condition $r_1=0$ implies
$r_2=0$ because $v_2^{(i)}$ must be proportional. Thus, either
logarithmic terms are present in both local solutions, or are not
present at all.

Moreover, condition $x=n\in\mathbb{N}$ is a necessary condition for
the confluent Heun function to be a polynomial of degree $(n-1)$. It
coincides with the condition
\begin{equation}
  \label{eq:9}
  \delta_{n-1}=: \frac{\delta}{\alpha}+\frac{1}{2}\left( \beta+\gamma \right)+n=0\text{,}
\end{equation}
(see condition (1.5a) in Fiziev \cite{Fiziev:10::}); additionally, the
condition $r_i=0$ coincides with the condition $\Delta_{n-1}$ in
Fiziev \cite{Fiziev:10::}, and it guarantees that the confluent Heun
function $v(y)=Q_{n-1}(y)$ is a polynomial of degree $(n-1)$.

It is now demonstrated that, assuming $x=n\in\mathbb{N}$ and $r_i=0$,
for $i=0$ or $i=1$, Eq.~\eqref{eq:Heun} has another independent entire
solution. To wit, the dependent variable in~\eqref{eq:Heun} is
changed, setting $v(y):= \exp \left[-4\lambda^2y \right]w(y)$. Again,
this yields the confluent Heun equation of the form~\eqref{eq:Heun}, but with
new parameters $(\widetilde\alpha,\widetilde\beta, \widetilde\gamma,
\widetilde\delta,\widetilde\eta)
=(-\alpha,\beta,\gamma,\delta,\eta)$. Consequently, condition (1.5a)
from Fiziev \cite{Fiziev:10::} can be alternatively expressed as
\begin{equation}
  \label{eq:10}
  \frac{\widetilde\delta}{\widetilde\alpha}+
  \frac{1}{2}\left(\widetilde \beta+\widetilde\gamma \right)+n+1=0
\end{equation}
and it is automatically fulfilled. Moreover, it can be shown that
condition $\Delta_{n+1}$ for the transformed equation coincides with
$r_i=0$. Thus, the transformed equation admits a polynomial solution
$w(y)=R_{n}(y)$ of degree $n$. To rephrase, with the given assumption,
eigenvalues $x=n\in\mathbb{N}$ are doubly degenerate. These are the
classical Judd eigenstates \cite{Judd:79::,Kus:86::}.  Both eigenstates are expressed
in terms of polynomial and exponential functions. In~\cite{Reik:82::} these solution are expressed as truncated Neumann series, but using transcendent functions, i.e., Bessel functions, is
unnecessary here.

The above analysis of integer values of $x$ is incomplete. Assuming
that $x=n\in\mathbb{N}$ belongs to the spectrum, and $v(y)$ is the
corresponding eigenvalue given by an entire solution of
Eq.~\eqref{eq:Heun}, an expansion of $v(y)$ at singular point $y_i$
may have the form proportional to local solutions $v_{1}^{(i)}(y)$ for
$i=0,1$, which correspond to the non-zero exponents.  It seems that
this possibility was overlooked in earlier investigations.  The only place where it is mentioned that for $x = n$ there are other more complicated solutions is paper~\cite{Reik:82::}. However, only two numerical values of the
parameters are given without any further discussion.

Let us note that, for $x = n$, solutions $v_{1}^{(i)}(y)$ are locally holomorphic without additional conditions
(such as vanishing of the logarithmic term). Thus, in some sense, it is
more natural to assume that local expansions of $v(y)$ at singular points are proportional to $v_{1}^{(i)}(y)$  than that they are proportional to $v_{2}^{(i)}(y)$ 
which corresponds to already mentioned Judd states. In this case, local
solutions $v_{1}^{(0)}(y)$ and $v_{1}^{(1)}(y)$, as two expansions of
the same entire function $v(y)$, must be proportional to each other in
their common domain of definition. So,
\begin{equation}
  \label{eq:3}
  v_{1}^{(0)}(y)=\alpha v_{1}^{(1)}(y)\text{,}
\end{equation}
for a certain constant $\alpha$, and for all $y$ belonging to the
common part of the domains of definition of considered
solutions. Denoting $v_{(0)}(y):=v_{1}^{(0)}(y) $ and
$v_{(1)}(y):=v_{1}^{(1)}(y)$ the condition shown in~\eqref{eq:3} is
expressed in a Wronskian form,
\begin{equation}
  \label{eq:4}
  w(y):=\det \begin{bmatrix}
    v_{(0)}(y) &  v_{(1)}(y) \\ 
    v_{(0)}'(y) &  v_{(1)}'(y) 
  \end{bmatrix}=0\text{.}
\end{equation}
As previously mentioned, if the Wronskian vanishes at one point, then
it vanishes identically. Thus, in this case, the condition that $x=n$
belongs to the spectrum can be written in the form $w(1/2)=0$. This
provides a condition on the parameters $n$, and $p=(\mu,\lambda)$ of
the problem. Unlike the classical Judd solutions, the series does not
terminate, and solutions $v_{(0)}(y)$ and $v_{(1)}(y)$ can only be found
recursively. Moreover, these solutions are given as expansions at
different points.  This problem is now solved explicitly using Heun
confluent functions. For the properties and more information on those functions 
see Ronveaux~\cite[Section~B]{Ronveaux:95::}. It is found that
\begin{equation*}
  \begin{split}
    & v_{(1)}(y):= (y-1)^{-\gamma}\operatorname{HeunC}(c_1,1-y)\text{,} \\
    &v_{(0)}(y):=
    y^{-\beta}(y-1)^{-\gamma}\operatorname{HeunC}(c_0,y)\text{,}\\
    & c_1=\left(-\alpha, -\gamma, \beta, -\delta, \delta+\eta
    \right)\text{,} \quad c_0=\left(\alpha, -\beta, -\gamma, \delta,
      \eta \right)\text{.}
  \end{split}
\end{equation*}
Denoting $H_i(z)=\operatorname{HeunC}(c_i,z) $, and $h_i=H_i(1/2)$,
$h_i'=H_i'(1/2)$ for $i=0,1$, it is found that the Wronskian $w(1/2)$
is proportional to the following function
\begin{equation}
  \label{eq:7}
  F_n(\lambda,\mu) := h_0h'_1+h_1 \left(2n  h_0 +h_0' \right)\text{.} 
\end{equation}
In this case, parameters $c_0$ and $c_1$ are expressed in terms of $n$
and $p=(\lambda, \mu)$.

For each $n\in\mathbb{N}$, the following set is defined
\begin{equation}
  \label{eq:Fn}
  \mathscr{F}_n:=  \left\{ (\lambda,\mu)\in\mathbb{R}^2\,|\, 
    F_n(\lambda,\mu)=0 \right\}\text{.}
\end{equation}
This is the second component of spectral set $\mathscr{S}_n$.

To summarize, the most significant result of this Letter is that the
Judd solutions are only a special, finite subset of all eigenstates
with integer $x=E+\lambda^2$. Here presented are hitherto unknown closed form 
conditions that also allow the system to have such eigenstates. In the
Juddian case, the conditions are polynomial in $\mu$ and $\lambda$,
and hence provide a finite number of solutions (e.g. $\mu$ in terms
of $\lambda$). In contrast, the newly-discovered conditions are seemingly 
transcendental and yield,  as numerical investigations show,  infinitely many curves in the parameter plane
$(\lambda,\mu)$, so there are infinitely many choices of $\mu$ for a
given $\lambda$. Together with considerations for non-integer $x$,
this analysis fully describes in a uniform framework the spectrum of the Rabi model,
i.e. provides eigenenergies, eigenstates and the corresponding
restrictions on all the other parameters.

Finally, we wish to comment on the questions of solvability and integrability
of the Rabi model touched upon in \cite{Braak:11::}. It should first be made
clear that the frequently used word
``analytic(al)'' has two meanings in this context. First, in connection with entire
functions, meaning complex differentiable, as used by Bargmann himself in the
title of his article \cite{Bargmann:61::}. Second and less rigorous, in context
of solvability, where it usually means ``expressible by elementary functions''
or ``explicit formulae''. We avoid the latter meaning and speak of solvability
in that case, for example in quadratures or in terms of Liouvillian functions
\cite{Zoladek:98::}. It should thus be understood that although the
eigenfunctions here are analytic in the first sense, they do not, in general
amount to what one would call explicit solutions. They can be written as Heun
functions, which was discovered in \cite{Maciejewski:12::x}, but one
has to understand the jump in complexity from the $_2F_1$ hypergeometric function to
$\mathrm{HeunC}(z)$, which is given by a series $\sum_n c_n z^n$, whose coefficients have to
be determined recursively, i.e., $c_n$ are not given as explicit functions of
the index $n$ (except for degenerate cases, see the next paragraph). Let us
recall here, that for {\em any} linear system, a solution
around a regular singular point can be constructed by the Frobenius method, as
such series so one cannot consider providing such a series a proof of
solvability, for then {\em all} such systems would be solvable. With symbolic packages
such as Maple now offering the Heun functions, the approach of
\cite{Maciejewski:12::x} or \cite{Zhong:13::} can be said to at least provide explicit
formulae but the $G_{\pm}$ functions of \cite{Braak:11::} are just
Frobenius series which do not indicate the solvability let alone integrability
claimed therein.

Since there is the distinguished subset $\mathscr{J}_n$ of the spectrum, 
the Rabi model can be called quasi-solvable, as the Heun functions reduce
to polynomials in the Juddian solutions -- for a part of the spectrum we know
both the exact eigenenergies and the eigenstates explicitly in terms of
elementary functions. However, since the $\mathbb{Z}_2$ symmetry
seems to have no direct impact on the solvability, we feel that the parity
operator is not enough to speak of quantum integrability for this system, for
which another commuting operator would be required, providing more than a
finite discrete symmetry.

This research was supported by grant No.  DEC-2011/ 02/A/ST1/00208 of
National Science Centre of Poland.

%
\def\polhk#1{\setbox0=\hbox{#1}{\ooalign{\hidewidth
  \lower1.5ex\hbox{`}\hidewidth\crcr\unhbox0}}} \def\cprime{$'$}
  \def\cydot{\leavevmode\raise.4ex\hbox{.}} \def\cprime{$'$}

\end{document}